Physical Review Letters

Title: Search for Charge 1/3 Third Generation Leptoquarks in pbarp
Collisions at sqrt(s)=1.8 TeV
Authors: B. Abbott et al. (D0 Collaboration)
Corresponding author: David Hedin
Email: hedin@niu.edu
Address: Department of Physics
Northern Illinois University
DeKalb, IL 60115
Phone number: 815-753-6483
FAX number: 815-753-8565

Article type: Letter
Experiment or theory: Experiment

Suggested section: L-1 Elementary particles and fields

Suggested principal PACS No.: 14.80.-j
Additional PACS No(s).: 13.85.Rm    

Expected figures: PostScript figures included in this electronic submission
Submitted by: hedin@niu.edu 

Submission of this manuscript implies acceptance by the authors
of the established procedures for selecting manuscripts for publication.
It is understood that this manuscript is original work and is not
being considered for publication elsewhere.

\documentstyle[prl,aps,floats,psfig]{revtex}

\def\met{\mbox{${\hbox{$E$\kern-0.6em\lower-.1ex\hbox{/}}}_T$}} 
\def\D0{D\O}                            
\def\d0draft{}
\def\err#1#2#3 {{\it Erratum} {\bf#1},{\ #2} (19#3)}
\def\ib#1#2#3 {{\it ibid.} {\bf#1},{\ #2} (19#3)}
\def\nc#1#2#3 {Nuovo Cim. {\bf#1} ,#2(19#3)}
\def\nim#1#2#3 {Nucl. Instr. Meth. {\bf#1},{\ #2} (19#3)}
\def\np#1#2#3 {Nucl. Phys. {\bf#1},{\ #2} (19#3)}
\def\pl#1#2#3 {Phys. Lett. {\bf#1},{\ #2} (19#3)}
\def\prev#1#2#3 {Phys. Rev. {\bf#1},{\ #2} (19#3)}
\def\prl#1#2#3 {Phys. Rev. Lett. {\bf#1},{\ #2} (19#3)}
\def\rmp#1#2#3 {Rev. Mod. Phys. {\bf#1},{\ #2} (19#3)}
\def\zp#1#2#3 {Zeit. Phys. {\bf#1},{\ #2} (19#3)}
\def\pp{$p\bar{p}$}

\begin{document}


%
%
\title{Search for Charge 1/3 Third Generation
Leptoquarks in \pp ~Collisions at $\sqrt{s} =$1.8 TeV}
\date{submitted March 10, 1998}

\author{                                                                      
B.~Abbott,$^{31}$                                                             
M.~Abolins,$^{27}$                                                            
B.S.~Acharya,$^{46}$                                                          
I.~Adam,$^{12}$                                                               
D.L.~Adams,$^{40}$                                                            
M.~Adams,$^{17}$                                                              
S.~Ahn,$^{14}$                                                                
H.~Aihara,$^{23}$                                                             
G.A.~Alves,$^{10}$                                                            
N.~Amos,$^{26}$
% ...
\\                                                                            
\vskip 0.50cm                                                                 
\centerline{(D\O\ Collaboration)}                                             
\vskip 0.50cm                                                                 
}                                                                             
\address{                                                                     
\centerline{$^{1}$Universidad de los Andes, Bogot\'{a}, Colombia}             
\centerline{$^{2}$University of Arizona, Tucson, Arizona 85721}               
\centerline{$^{3}$Boston University, Boston, Massachusetts 02215}             
\centerline{$^{4}$Brookhaven National Laboratory, Upton, New York 11973}      
\centerline{$^{5}$Brown University, Providence, Rhode Island 02912}           
\centerline{$^{6}$Universidad de Buenos Aires, Buenos Aires, Argentina}       
\centerline{$^{7}$University of California, Davis, California 95616}          
\centerline{$^{8}$University of California, Irvine, California 92697}         
\centerline{$^{9}$University of California, Riverside, California 92521}      
\centerline{$^{10}$LAFEX, Centro Brasileiro de Pesquisas F{\'\i}sicas,        
                  Rio de Janeiro, Brazil}                                     
\centerline{$^{11}$CINVESTAV, Mexico City, Mexico}                            
\centerline{$^{12}$Columbia University, New York, New York 10027}             
\centerline{$^{13}$Delhi University, Delhi, India 110007}                     
\centerline{$^{14}$Fermi National Accelerator Laboratory, Batavia,            
                   Illinois 60510}                                            
\centerline{$^{15}$Florida State University, Tallahassee, Florida 32306}      
\centerline{$^{16}$University of Hawaii, Honolulu, Hawaii 96822}              
\centerline{$^{17}$University of Illinois at Chicago, Chicago,                
                   Illinois 60607}                                            
\centerline{$^{18}$Indiana University, Bloomington, Indiana 47405}            
\centerline{$^{19}$Iowa State University, Ames, Iowa 50011}                   
\centerline{$^{20}$Korea University, Seoul, Korea}                            
\centerline{$^{21}$Kyungsung University, Pusan, Korea}                        
\centerline{$^{22}$Institute of Nuclear Physics, Krak\'ow, Poland}            
\centerline{$^{23}$Lawrence Berkeley National Laboratory and University of    
                   California, Berkeley, California 94720}                    
\centerline{$^{24}$Louisiana Tech University, Ruston, Louisiana 71272}        
\centerline{$^{25}$University of Maryland, College Park, Maryland 20742}      
\centerline{$^{26}$University of Michigan, Ann Arbor, Michigan 48109}         
\centerline{$^{27}$Michigan State University, East Lansing, Michigan 48824}   
\centerline{$^{28}$Institute for Theoretical and Experimental Physics,        
                   Moscow, Russia}                                            
\centerline{$^{29}$Moscow State University, Moscow, Russia}                   
\centerline{$^{30}$University of Nebraska, Lincoln, Nebraska 68588}           
\centerline{$^{31}$New York University, New York, New York 10003}             
\centerline{$^{32}$Northeastern University, Boston, Massachusetts 02115}      
\centerline{$^{33}$Northern Illinois University, DeKalb, Illinois 60115}      
\centerline{$^{34}$Northwestern University, Evanston, Illinois 60208}         
\centerline{$^{35}$University of Notre Dame, Notre Dame, Indiana 46556}       
\centerline{$^{36}$University of Oklahoma, Norman, Oklahoma 73019}            
\centerline{$^{37}$University of Panjab, Chandigarh 16-00-14, India}          
\centerline{$^{38}$Institute for High Energy Physics, Protvino 142284,        
                   Russia}                                                    
\centerline{$^{39}$Purdue University, West Lafayette, Indiana 47907}          
\centerline{$^{40}$Rice University, Houston, Texas 77005}                     
\centerline{$^{41}$Universidade do Estado do Rio de Janeiro, Brazil}          
\centerline{$^{42}$University of Rochester, Rochester, New York 14627}        
\centerline{$^{43}$CEA, DAPNIA/Service de Physique des Particules,            
                   CE-SACLAY, Gif-sur-Yvette, France}                         
\centerline{$^{44}$Seoul National University, Seoul, Korea}                   
\centerline{$^{45}$State University of New York, Stony Brook,                 
                   New York 11794}                                            
\centerline{$^{46}$Tata Institute of Fundamental Research,                    
                   Colaba, Mumbai 400005, India}                              
\centerline{$^{47}$University of Texas, Arlington, Texas 76019}               
\centerline{$^{48}$Texas A\&M University, College Station, Texas 77843}       
}                                                                             
\maketitle
\begin{abstract}

We report on  a search for charge 1/3 third
generation leptoquarks (LQ) produced in $p\bar{p}$ collisions at $\sqrt s = 1.8$ TeV
using the \D0 detector at Fermilab. Third generation
leptoquarks are assumed to be produced in pairs and 
to decay to a tau neutrino and a $b$ quark with branching fraction
${B}$. We place
upper limits on $\sigma (p\bar{p} \rightarrow LQ~\overline{LQ})\cdot{B}^2$
as a function of the leptoquark mass $M_{LQ}$. 
Assuming ${B} =1$, we exclude at the 95\% confidence level
third generation scalar leptoquarks with $M_{LQ}<94$ GeV/$c^2$,
and  third generation vector leptoquarks with 
$M_{LQ}<216$ GeV/$c^2$ ($M_{LQ}<148$ GeV/$c^2$) assuming
Yang-Mills (anomalous) coupling.

\end{abstract}
\twocolumn
%

Leptoquarks (LQ) are bosons predicted in
many extensions to the standard model \cite{th1}. They carry
both lepton and color quantum numbers, couple to
leptons and quarks, and decay via
$LQ\rightarrow l+q$. To satisfy experimental constraints
on flavor changing neutral currents, leptoquarks of mass accessible
to current collider experiments are constrained to couple to 
only one generation of leptons and quarks\cite{th2}.
Therefore, only leptoquarks which couple within
a single generation are considered here.

This Letter reports the results of a search for charge 1/3
third generation leptoquarks produced in
\pp ~collisions at $\sqrt s = 1.8$ TeV. We assume that
leptoquarks are produced in pairs by QCD processes such as 
$p\bar{p} \rightarrow g \rightarrow LQ~ \overline{LQ}+ X$.
This process dominates
over other production mechanisms which depend
on the unknown leptoquark-lepton-quark coupling
$\lambda$ under the standard condition $\lambda \le \sqrt{4\pi}\alpha_{EM}$.
We search for the decay signature where both leptoquarks decay via
$LQ\rightarrow \nu_\tau + b$ resulting in a 
$\nu_\tau\bar{\nu}_\tau b\bar{b}$ final state. 
For a leptoquark mass ($M_{LQ}$) smaller than the mass of
the top quark ($m_t$), the decay $LQ\rightarrow \tau +t$
is forbidden, and the branching fraction for the $\nu_\tau b$ mode, ${B}$, 
is unity\cite{beta}.  For $M_{LQ}>m_t$, phase space factors suppress 
$\tau t$ decays relative to the $\nu_\tau b$ channel.
In this paper we give
limits on the pair production cross section times ${B} ^2$ ($\sigma \cdot B^2$)
for $M_{LQ}$ between 50 GeV/$c^2$ and 300 GeV/$c^2$.
Limits on the cross section are used to set limits on
the third generation leptoquark mass for 
scalar and vector leptoquarks. 
Previous limits from the LEP $e^+e^-$ collider exclude all third generation
leptoquarks with masses below 45 GeV/$c^2$\cite{lep},
while the CDF Collaboration has set limits for pair produced charge 2/3 or 4/3
third generation leptoquarks decaying via $LQ\rightarrow \tau + b$\cite{cdf}.  

The data used for this analysis were collected by the \D0
detector\cite{dzero} operating at the Fermilab Tevatron Collider
during the 1993--1996 period. 
The \D0 detector
is composed of three major systems: an inner detector for
tracking charged particles, a uranium/liquid argon calorimeter for
measuring electromagnetic and hadronic showers, and a muon spectrometer
consisting of a magnetized iron toroid and three layers of drift tubes.
The detector measures jets with an energy resolution of approximately   
$\sigma /E = 0.8/\sqrt E$ ($E$ in GeV) and muons with a momentum
resolution of 
$\sigma /p=[(\frac{0.18(p-2)}{p})^2 + (0.003 p)^2]^{1/2}$  ($p$ in GeV$/c$).
Missing transverse energy (\met) is determined by
summing the calorimeter and muon transverse energies, and is measured with
a resolution of $\sigma$ = 1.08 GeV + 0.019$\cdot (\Sigma |E_T|)$ ($E_T$ in GeV).

The decay of a $b$ quark is indicated by the presence of a muon
associated with a jet. We use three triggers to collect
candidate leptoquark events, each requiring one or two muons\cite{level1}.
A dimuon trigger required two muons 
with transverse momentum $p_T^\mu>3.0$ GeV/$c$. One single muon
trigger required a muon with $p_T^\mu> 1.0$ GeV/$c$ and a 
jet  with $E_T^j > 10$ GeV. 
The other single muon trigger required a muon
with $p_T^\mu> 10$ GeV/$c$ in the trigger and $p_T^\mu> 15$ GeV/$c$ during offline
analysis, and a jet  with $E_T^j> 15$ GeV.
Integrated luminosities of 60.1 pb$^{-1}$, 19.5 pb$^{-1}$,
and 92.4 pb$^{-1}$ respectively were collected
using these three triggers.   

The offline analysis uses muons   
in the pseudorapidity range $|\eta_\mu|$ $<$  1.0
with $p_T^\mu>3.5$ GeV/$c$. The
muon trajectories  are required to be consistent 
with the reconstructed vertex position and have associated 
energy  in the calorimeter.
For events from either single muon trigger additional
requirements are imposed:  
 the presence of hits in all three muon detector layers,
a matching track in the central
detector, and a good fit\cite{huehn} when these elements are combined.
For the dimuon trigger events, at least one of the two muons must
satisfy each of these additional requirements.
Jets are reconstructed using only calorimeter energy with
a cone algorithm of radius
$R=0.7$ where $R= \sqrt{(\Delta\phi)^2+(\Delta\eta)^2}$ about
the jet's centroid and $\phi$ is the azimuthal angle. Each jet is required to
have $E_T^j>10$ GeV and to satisfy reconstruction quality criteria \cite{susy}.
 For the dimuon trigger, events with
dimuon invariant mass greater than 8 GeV/$c^2$ are selected 
to eliminate backgrounds from low mass resonances and
each muon must be
associated with a different jet with  $\Delta R_{\mu - jet}<0.5$
to increase $b$ quark purity \cite{fein}.  For the single muon triggers, we require a
jet associated with the muon with the same $\Delta R$ 
 requirement plus an additional jet with $E_T^j>25$ GeV 
and $|\eta_j|<1.5$.

We use \met~to identify neutrinos.
To help eliminate events with poorly measured \met,
we reject events where the azimuthal
angular separation, $\Delta\phi$, between the missing energy  and
the nearest jet is less than 0.7 radians.  
 In some
events, the source of the measured \met~is not 
mismeasured jets, but rather noise in 
individual calorimeter cells or from cells
activated by the background generated by the Main Ring 
accelerator, which passes through the calorimeter.
Such events are removed if the vector sum of the transverse energies
in the jets and muons is consistent with zero.

\begin{figure}\vbox{
\centerline{
\psfig{figure=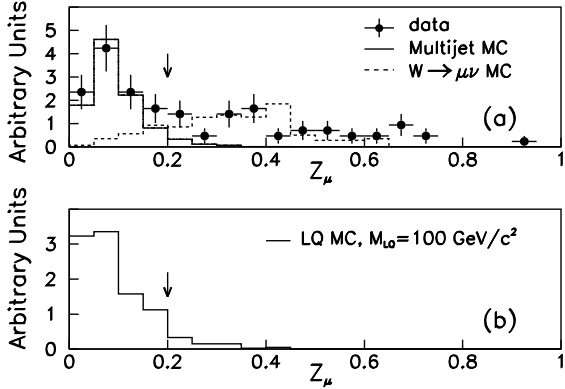,width=3.3in}}
\caption{ The $Z_\mu$ distribution  for  events
with \met$>30$ GeV
and $\Delta \phi>0.6$ from (a) data  
compared to MC events from  multiple
jet (solid) and $W\rightarrow \mu\nu$ (dashed) processes   
and (b) leptoquark events with $M_{LQ}=100$ GeV/$c^2$. The
arrow indicates the requirement $Z_\mu < 0.20$ used in this
analysis.}
\label{fig:zmu}
}
\end{figure}

Data from the two single muon triggers
with large \met~have contributions from $W+\ge 2$ jet events
where $W\rightarrow \mu\nu$ and the muon overlaps a jet. 
We use two variables, $Z_\mu$ and $F_\mu$,  for events passing those
triggers to reduce the $W$ boson acceptance. 
We define $Z_\mu \equiv p_T^\mu/H_T^{\mu j}$, where $H_T^{\mu j}$ is
the scalar sum of the $E_T$ of the jets and muons in the event.
In Fig.~\ref{fig:zmu}  we compare the
$Z_\mu$ distributions of  data events that
pass the low-$p_T$ single muon trigger to Monte Carlo (MC) samples \cite{monte}
satisfying a simulation of the trigger. There, for illustrative purposes,
we select events with \met$>30$ GeV and $\Delta \phi>0.6$ so
that the data shown have roughly equal contributions
from $W$ boson and hadronic multijet events. Also shown are 
MC distributions of equal numbers of multijet
events (which are $b$ quark-dominated due to their muon content)
and $W\rightarrow \mu\nu$ events.   
The data (normalized to the total number
of MC events) are consistent with the sum of these sources.
Also shown is the same distribution for a leptoquark MC sample
with $M_{LQ}=100$ GeV/$c^2$ satisfying the same
criteria. Since the leptoquark $Z_\mu$
distribution is determined by $b$ quark decay kinematics, it is similar to 
the multijet distribution. Requiring $Z_\mu < 0.20$ eliminates about 90\% of
the remaining $W\rightarrow \mu\nu$ background while maintaining 
a signal efficiency of 82\% for $M_{LQ}=100$ GeV/$c^2$.
The second variable, $F_\mu$,  is the ratio of calorimeter energy
within a cone of 0.4 centered on the muon's direction to that within 
a cone of 0.6: $F_\mu \equiv  E(R_\mu\leq 0.4)/E(R_\mu\leq 0.6)$. 
Most hadronic energy in higher $E_T$ direct b quark decays is spatially close
to the muon.
Requiring $F_\mu > 0.80$ removes about 84\% of
the $W\rightarrow \mu\nu$ background with a leptoquark signal efficiency of 82\%.
MC studies indicate that  $Z_\mu$ and $F_\mu$ have  little
correlation.

 The data from the high-$p_T$ muon trigger include a significant
contribution from top quark pair production ($t\bar{t}$) events. The
scalar sum of jet $E_T$, $H_T^j$, was used to identify the
top quark in  Ref. \cite{top1}. In this analysis, we require $H_T^{\mu j}<240$ GeV
for those events satisfying the high-$p_T$ single
muon trigger to reduce the $t\overline{t}$
contribution. Similarly, since the low-$p_T$ single muon trigger
has a larger contribution from multijet events, 
we reject events with six or more jets ($E_T^j>10$ GeV) for that trigger. 

\begin{figure}\vbox{
\centerline{
\psfig{figure=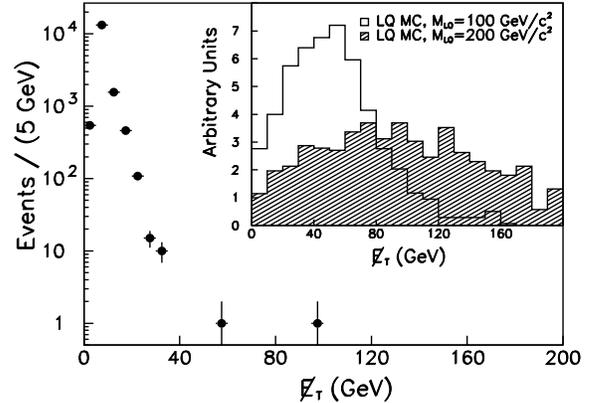,width=3.3in}}
\caption{ The distributions of \met~for data events
after all other selection requirements have been applied.
Shown in the insert are \met~distributions for MC leptoquark
events with leptoquark mass of 100  GeV/$c^2$
 and  200  GeV/$c^2$.}
\label{fig:met}
}
\end{figure}

The resulting \met~distribution for data from
all three triggers after all selection
criteria have been applied 
is given in Fig.~\ref{fig:met}. 
Also shown are the \met~distributions for MC events with 
$M_{LQ} =$ 100 and 200 GeV/$c^2$.
Requiring \met$>35$ GeV leaves two events.
 
We  consider background contributions from $t\bar{t}$,
intermediate vector boson, and multijet production of $b\bar{b}$ and
$c\bar{c}$. Top quark events have multiple $b$ quarks and \met,
but with additional, energetic jets. We use MC and
our measured $t\bar{t}$ production cross section \cite{top2}
to estimate that there are $1.4\pm 0.5$ $t\bar{t}$  
events in our sample. Intermediate vector boson events 
have \met~from $W\rightarrow l\nu$ or $Z\rightarrow \nu\bar{\nu}$ and
muons near jets mimicking $b$ quark decays when either a prompt muon 
overlaps a jet or a jet fragments into
a muon via $c$ quark or a $\pi$/$K$ decay.  
Using our measured $W$ and $Z$ boson production cross sections \cite{wz} 
yields $1.0 \pm 0.4$ $W$ boson events
and $0.1\pm 0.1$ $Z$ boson events in this sample. 
Hadronically-produced $b\bar{b}$ and $c\bar{c}$ events do not have energetic neutrinos
and are effectively eliminated by the \met~and $\Delta\phi$ cuts.
Estimates of their contribution using data and MC
are consistent with zero, and we conservatively assume this
in our limit calculation.
Therefore, the total background is estimated to be $2.5\pm 0.6$ events.

\begin{table}
\caption{Third generation scalar leptoquark acceptances for
the three trigger channels and     
95\% C.L. limits on $\sigma \cdot B^2$ for different $M_{LQ}$. }
\begin{tabular}{c c c c c}
& \multicolumn{3}{c}{Acceptance ($\times 10^{-3}$)}  \\
LQ mass  & dimuon  & one muon  & one muon & $\sigma \cdot B^2$    \\ 
(GeV/$c^2$)   &   &  low-$p_T$  &  high-$p_T$  & limit (pb)\\ 
\hline
~50 & 0.04& ~0.95 & 0.14 & 144~~ \\
~60 & 0.08 & ~2.4~ & 0.35 & ~59~~ \\
~80 & 0.29 & ~6.0~ & 0.93 & ~22.6 \\
100 & 0.40 & ~9.7~ & 1.9~& ~12.7  \\
125 & 0.59 & 14~~~ & 2.8~ & ~~8.9 \\
150 & 1.2~ & 18~~~ & 3.2~ & ~~7.0 \\
200 & 1.3~ & 24~~~ & 3.1~ & ~~6.0 \\
250 & 1.9~ & 27~~~ & 3.7~ & ~~5.1 \\
300 & 1.8~ & 31~~~ & 2.2~ & ~~5.4 \\
\end{tabular}
\label{tab:table2}
\end{table}

We calculate the detection efficiency for scalar leptoquark signals using MC
acceptances multiplied by muon trigger and
reconstruction efficiencies obtained from data samples collected using
test triggers. The acceptances for different leptoquark masses
are summarized in Table~\ref{tab:table2}. The use of a 
muon to tag $b$ quark decays
limits the acceptance to values under 3.5\%.
Factors contributing to this limited acceptance 
for the low-$p_T$ single muon channel 
with $M_{LQ}=100$ GeV/$c^2$
include the muon branching fraction (0.35), muon and
jet kinematic requirements (0.35), 
and muon trigger and reconstruction efficiency
(0.25).
The requirements used to reduce the
background (\met, $\Delta\phi$, $Z_\mu$, $F_\mu$)
retain  $\approx 40$\% of the leptoquark signal.

We combine the three trigger channels to set limits.
Errors on the acceptance are shown
in Table~\ref{tab:table3}. Errors on trigger and
reconstruction efficiencies are due to the
statistical errors of the data used to
calculate their values. Muon momentum
resolution and jet energy scale errors are obtained
from data and their impact on the acceptance is
determined using MC with $M_{LQ}=100$ GeV/$c^2$. The error on the $F_\mu$ cut efficiency is
estimated by comparing data events without 
\met~requirements to MC multijet events. The three trigger channels have
different systematic errors since their selection criteria
and average muon $p_T$ differ, but  most errors are
correlated. The total systematic error, including
correlations and MC statistics, on the combined acceptance varies 
between 12.5\% and 13.6\% for different leptoquark masses. 

\begin{table}
\caption{Summary of systematic errors in terms
of \% error on the acceptance.}
\begin{tabular}{c c c c}
 Channel & dimuon  & one muon   & one muon    \\ 
  &   & low-$p_T$  & high-$p_T$   \\ 
\hline
trigger   &~6.5 &~5.1  & ~5.1 \\
reconstruction   &~5.1 &~5.7  &~4.2 \\
muon momentum resolution &~1~~ &~1~~  & 10~~\\
jet energy scale   &~4~~ & ~4~~ & ~2~~ \\
$F_\mu$ cut &  NA &10~~& ~6~~  \\
$b\rightarrow \mu$ fraction & 12~~ & ~6~~ & ~6~~ \\
\end{tabular}
\label{tab:table3}
\end{table}

\begin{figure}\vbox{
\centerline{
\psfig{figure=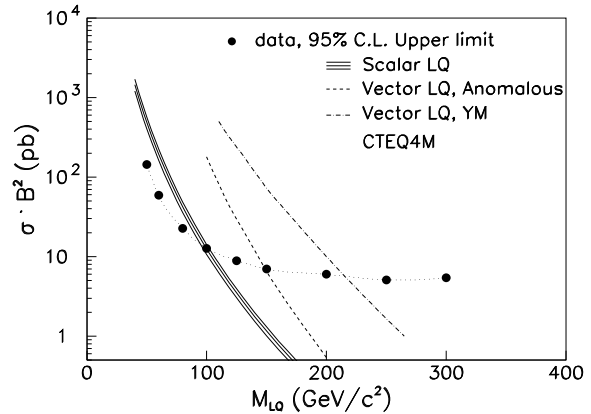,width=3.3in}}
\caption{ The 95\% C.L. limit  on $\sigma \cdot B^2$
($\bullet$) compared to theoretical predictions. 
The prediction for scalar leptoquarks (solid) 
use  $\mu = 2M_{LQ}, ~M_{LQ}$ and $M_{LQ}/2$, while
vector leptoquarks with  minimal anomalous (dashed) or
Yang-Mills (dot-dashed) coupling use  $\mu = M_{LQ}$.}
\label{fig:cl}
}
\end{figure}

The 95\% confidence level (C.L.) upper limits on  
$\sigma \cdot B^2$ include the
systematic acceptance uncertainty and a 5.3\% uncertainty in
the integrated luminosity. The resulting upper limits  
 for scalar leptoquark pair production as a 
function of leptoquark mass are given in Table I and shown in 
Fig.~\ref{fig:cl}. Also
shown in Fig.~\ref{fig:cl} are theoretical cross sections
for the production of scalar and vector leptoquarks.
The calculation of the scalar leptoquark cross section includes next-to-leading
order diagrams and uses CTEQ4M parton distribution functions\cite{NLO}.
The theory band shown in the figure is determined using
a renormalization factor of $\mu = M_{LQ}$ for the central
value and $\mu = 2M_{LQ}$ and $M_{LQ}/2$ for the lower and
upper bounds, respectively.  
The intersection of our limit curve with the lower
edge of the theory band is at 94 GeV/$c^2$.
This is our
95\% C.L. lower limit on the mass of a charge 1/3 third
generation scalar leptoquark (taking ${B} = 1$ since $M_{LQ}< m_t$).

Similarly, we set limits for the mass of vector leptoquarks \cite{veca}.
The vector leptoquark cross section has been calculated in the
leading order approximation using CTEQ4M parton distribution functions
and  $\mu=M_{LQ}$ \cite{vector}. Vector leptoquarks are assumed to be
either fundamental gauge bosons with Yang-Mills coupling or
composite particles with anomalous coupling. For 
Yang-Mills type coupling, a mass limit of
216 GeV/$c^2$ is obtained for ${B} = 1$. 
If $M_{LQ}> m_t$ the $\tau t$ mode is allowed.  We consider the case in which the
branching fraction to $\nu_\tau b$ and $\tau t$ each would be 0.5 if the fermion
masses could be neglected relative to $M_{LQ}$.  Taking into account phase space
suppression factors \cite{rizzo}, we determine that $M_{LQ}> 209$ GeV/$c^2$
for Yang-Mills type vector leptoquarks for this $B < 1$ case.
For anomalous coupling, we choose 
the coupling which yields the minimum pair production cross section.
The intersection of our limit on $\sigma \cdot B^2$ with the theory
curve gives  $M_{LQ}>148$ GeV/$c^2$ for minimal anomalous vector coupling.

In conclusion, we observe two events consistent with the
final state $\nu\bar{\nu} b\bar{b}$ compared to an expected $2.5\pm 0.6$
events from $t\bar{t}$ and $W$ and $Z$ boson production. We set limits
on the mass of a charge 1/3 scalar or vector leptoquark. This result is independent of the
coupling strength of a leptoquark to a third generation lepton
and quark. 

We thank the staffs at Fermilab and collaborating institutions for their
contributions to this work, and acknowledge support from the 
Department of Energy and National Science Foundation (U.S.A.),  
Commissariat  \` a L'Energie Atomique (France), 
State Committee for Science and Technology and Ministry for Atomic 
   Energy (Russia),
CAPES and CNPq (Brazil),
Departments of Atomic Energy and Science and Education (India),
Colciencias (Colombia),
CONACyT (Mexico),
Ministry of Education and KOSEF (Korea),
and CONICET and UBACyT (Argentina).


\begin{references}
\bibitem{beta} ${B}$ is defined as $1-\beta$, where $\beta$ is the leptoquark branching
fraction into a charged lepton and quark.
\bibitem{veca} For masses greater than 150 GeV/$c^2$, MC acceptances
for vector leptoquarks are found to be greater than or equal to
those for scalar leptoquarks with the same $M_{LQ}$.
\end{references}
\end{document}